\documentclass[12pt]{article}
\usepackage{epsfig}
\usepackage{latexsym}
\DeclareGraphicsExtensions{.eps.gz,.eps,.ps,.ps.gz}
\parskip=\itemsep               
\date{\empty}

\catcode`@=11

\def\@citex[#1]#2{\if@filesw\immediate\write\@auxout{\string\citation{#2}}\fi
  \def\@citea{}\@cite{\@for\@citeb:=#2\do
    {\@citea\def\@citea{,\penalty\@m}\@ifundefined
      {b@\@citeb}{{\bf ?}\@warning
       {Citation `\@citeb' on page \thepage \space undefined}}%
\hbox{\csname b@\@citeb\endcsname}}}{#1}}

\def\citer{\@ifnextchar [{\@tempswatrue\@citexr}{\@tempswafalse\@citexr[]}}

\def\@citexr[#1]#2{\if@filesw\immediate\write\@auxout{\string\citation{#2}}\fi
  \def\@citea{}\@cite{\@for\@citeb:=#2\do
    {\@citea\def\@citea{--\penalty\@m}\@ifundefined
       {b@\@citeb}{{\bf ?}\@warning
       {Citation `\@citeb' on page \thepage \space undefined}}%
\hbox{\csname b@\@citeb\endcsname}}}{#1}}
\catcode`@=12

\def\bo{{\raise.15ex\hbox{\large$\Box$}}}               
\def\face{{\raise.2ex\hbox{$\displaystyle \bigodot$}\mskip-2.2mu \llap {$\ddot
        \smile$}}}                                      

\def\leftrightarrowfill{$\mathsurround=0pt \mathord\leftarrow \mkern-6mu
        \cleaders\hbox{$\mkern-2mu \mathord- \mkern-2mu$}\hfill
        \mkern-6mu \mathord\rightarrow$}       
\def\dvec#1{\vbox{\ialign{##\crcr
        \leftrightarrowfill\crcr\noalign{\kern-1pt\nointerlineskip}
        $\hfil\displaystyle{#1}\hfil$\crcr}}}           

\def\beq{\begin{equation}}
\def\eeq{\end{equation}}

\def\beqx{\begin{displaymath}}
\def\eeqx{\end{displaymath}}

\def\beql{\begin{eqnarray}}
\def\eeql{\end{eqnarray}}
\def\NO{\nonumber}

\def\Journal#1#2#3#4{{#1} {\bf #2} (#4) #3}

\def\NPB{{\em Nucl. Phys.} B}
 
\def\PLB{{\em Phys. Lett.} B} 
\def\PRL{\em Phys. Rev. Lett.} 
\def\PRD{{\em Phys. Rev.} D} 
 
\def\PR{\em Phys. Rev.}

\def\PR{\em Phys. Rep.}

\newcommand{\lwig}{\mbox{\,\raisebox{.3ex}
    {$<$}$\!\!\!\!\!$\raisebox{-.9ex}{$\sim$}\,}}
\newcommand{\gwig}{\mbox{\,\raisebox{.3ex}
    {$>$}$\!\!\!\!\!$\raisebox{-.9ex}{$\sim$}}\,}

\newcommand{\ai}{{\overline{I}}} 
\newcommand{\iai}{I\overline{I}}

\newcommand{\ii}{{\rm i}} 
 
\newcommand{\xpr}{{x^\prime}}

\begin{document}
\title{
{\normalsize\rightline{DESY 98-201}\rightline{hep-lat/9903039}} 
\vskip 1cm 
      \bf  Confronting Instanton Perturbation Theory\\ 
      \bf with QCD Lattice Results 
       \vspace{11mm}}
\author{A. Ringwald and F. Schrempp\\[4mm] 
Deutsches Elektronen-Synchrotron DESY, Hamburg, Germany}
\begin{titlepage} 
  \maketitle
\begin{abstract}
We exploit a recent lattice investigation (UKQCD) on the topological structure 
of the (quenched) QCD vacuum, in order to  gain information on crucial
building blocks of instanton perturbation theory. A central motivation
is to further constrain our previous predictions of instanton-induced hard
scattering processes. First, we address the generic problem of extracting
quantitative information from cooled lattice data. We find  a new scaling
variable, interpreted as a ``cooling radius'', which allows to 
combine lattice data for a whole range of lattice spacings and cooling 
sweeps. This variable strongly helps to extract information on the uncooled
distributions of interest. After performing the continuum
extrapolation of the instanton size and instanton-anti-instanton
distance distributions, we find striking agreement with the theoretical
predictions from instanton-perturbation theory, for instanton sizes
$\lwig 0.5$ fm and distances $\gwig 0.5$ fm, respectively. These
results imply first direct support for the validity of the known valley
interaction between instantons and anti-instantons.   
\end{abstract} 
\thispagestyle{empty}
\end{titlepage}
\newpage \setcounter{page}{2}

{\bf 1.}
Non-abelian gauge theories like QCD are known to exhibit a rich vacuum 
structure. The latter includes topologically non-trivial
fluctuations of the gauge fields, carrying an integer topological
charge $Q$.  The simplest building blocks  of topological structure
are  instantons ($Q=+1$) and anti-instantons  ($Q=-1$) which are well-known
explicit solutions of the euclidean field equations in four
dimensions~\cite{bpst}.  

Instantons are widely believed to play an important r{\^o}le in various 
long-distance aspects~\cite{ssh} of QCD: 

First of all, they may provide a solution of the famous $U_A(1)$ 
problem~\cite{th} ($m_{\eta^\prime}\gg m_{\eta}$), with the corresponding 
pseudoscalar mass splitting related to the topological susceptibility 
in the pure gauge theory by the well-known Witten-Veneziano formula~\cite{wv}.
Moreover, a number of authors have attributed a strong
connection  of instantons with chiral symmetry breaking~\cite{dia,ssh}
as well as the hadron and glueball spectrum.

However, there are also very important short-distance
implications~\citer{bb,mrs2} of QCD instantons:

They are known to induce certain processes  
which violate chirality in accord with the general
axial-anomaly relation~\cite{th} and which are forbidden in conventional
perturbation theory. Of particular interest in this context is the
deep-inelastic scattering (DIS) regime. Here, 
hard in\-stan\-ton-induced processes may both be 
calculated~\citer{mrs1,mrs2} within instanton-per\-tur\-ba\-tion
theory  and possibly be detected
experimentally~\cite{rs,grs,rs2,dis97-phen}. Since an experimental  
discovery of instanton-induced events would clearly be of basic significance, 
our ongoing theoretical and phenomenological  study of the discovery 
potential of instanton-induced DIS events is well motivated.
 
On the other hand, there has been much recent activity in the lattice 
community to ``measure'' topological fluctuations in 
lattice simulations~\cite{lattice} of QCD. Being
independent of perturbation theory, such simulations provide
``snapshots'' of the QCD vacuum including all possible
non-perturbative features like instantons. They  may be exploited to 
provide crucial support for important prerequisites of calculations in DIS,
like the validity of instanton-perturbation theory and 
the dilute instanton-gas approximation for small instantons of size
$\rho \leq {\mathcal O}(0.3)$ fm. Along these lines, we were 
able~\cite{rs-pl} to translate the lattice constraints into a 
``fiducial'' kinematical region for our predictions of the
instanton-induced DIS cross-section based on instanton-perturbation theory. 

The purpose of the present letter is to present a more thorough
confrontation of predictions from instanton-perturbation theory with
recent lattice simulations. Specifically, we concentrate on recent
high-statistics results from the UKQCD collaboration~\cite{ukqcd,mike},
try to carefully discuss the effects of cooling and perform an
extrapolation to the continuum before drawing our conclusions. 

{\bf 2.}
Let us start by considering various important quantities
relevant for instanton ($I$)-induced scattering processes, that may both
be calculated in $I$-perturbation theory and measured in
lattice simulations. In $I$-perturbation theory one expands the path
integral for the generating functional of the 
Green's functions about the known, classical instanton solution
$A_\mu=A^{(I)}_\mu+\ldots$ instead of  
the trivial vacuum   $A^{(0)}_\mu=0$ (pQCD) and obtains a
corresponding set of modified Feynman rules. Like in conventional
pQCD, the strong gauge coupling $\alpha$ has to be small.  

The $I$-induced contribution to hard scattering processes is
typically encoded in the total cross-section~\cite{rs-pl,mrs2}
associated with an $I$-induced partonic subprocess
$q^\prime\,p \Rightarrow X$,  
\begin{eqnarray}
\label{cs}
\sigma^{(I)}(q^\prime\,p \Rightarrow X)&=&
      \int d^4 R\ 
      \int\limits_0^\infty d\rho
      \int\limits_0^\infty d\overline{\rho}
      \ D(\rho)\, D(\overline{\rho})
       \int dU\,
      {\rm e}^{{-\frac{4\pi}{\alpha}}\,
      \Omega\left(U,\frac{R^2}{\rho\overline{\rho}},
      \frac{\overline{\rho}}{\rho} \right)}\, \{\ldots\}\,\\      
&&    \times F(\sqrt{-q^{\prime\,2}}\,\rho)\,
      F(\sqrt{-q^{\prime\,2}}\,\overline{\rho})
      F(\sqrt{-p^2}\,\rho)\,F(\sqrt{-p^2}\,\overline{\rho})\,
      {\rm e}^{\ii\, (p+q^\prime)\cdot R}\ .\NO
\end{eqnarray}  
$\sigma^{(I)}$ involves integrations over
all $I\,[\overline{I}]$-``collective 
coordinates'',  including the $I\,[\overline{I}]$-sizes 
$\rho\,[\overline{\rho}]$, the $\iai$-distance\footnote{\label{iai}Both an
instanton and an anti-instanton enter here, 
since cross-sections result from taking the modulus squared of an
amplitude in the single $I$-background. Alternatively, one may
view the cross-section (\ref{cs}) as a discontinuity of the $q^\prime 
p$ forward elastic scattering amplitude in the
$\iai$-background~\cite{rs-pl}.} 4-vector $R_\mu$ and the 
$\iai$ relative colour orientation $U$.  
With each initial-state parton, there is an associated ``form
factor''~\cite{mrs1,rs-pl}, 
\begin{equation}
\label{formfac}
F(x)=x\,K_1(x)\ \left\{
\begin{array}{lcllcll}
&\sim& \sqrt{\pi/(2\,x)}\exp(-x) &{\rm \ for\ }&
x& \rightarrow +\infty;\NO\\[2mm]
&=& 1 &{\rm \ for\ }& x&=0;\\[2mm] 
&&{\rm oscillating} &{\rm\ for}&x& {\rm\ imaginary}.\NO\\
\end{array}\right.
\end{equation}

An important quantity entering Eq.~(\ref{cs}) as generic weight is
the (reduced) $I$-size distribution in the dilute gas approximation, 
$D(\rho)$. It is known in the 
framework of $I$-perturbation theory for small 
$\alpha(\mu_r)\ln(\rho\,\mu_r)$, where $\mu_r$ is the renormalization
scale. After its pioneering evaluation at 
1-loop~\cite{th}  for $N_c=2$ and its generalization~\cite{ber} to
arbitrary  $N_c$, it is meanwhile available~\cite{morretal} in 2-loop
renormalization-group (RG) invariant form,
i.e. $D^{-1}\,dD/d\ln(\mu_r)=\mathcal{O}(\alpha^2)$,
\begin{eqnarray}
D({\rho})&= &
\frac{d_{\overline{\rm MS}}}{\rho^5}
\left(\frac{2\pi}{\alpha_{\overline{\rm MS}}(\mu_r)}\right)^{2\,N_c} 
\exp{\left(-\frac{2\pi}{\alpha_{\overline{\rm MS}}(\mu_r)}\right)}(\rho\,
\mu_r)^{\beta_0+(\beta_1-4\,N_c\beta_0 )
\frac{\alpha_{\overline{\rm MS}}(\mu_r)}{4\pi}} ,
\label{dens}
\\
 d_{\overline{\rm MS}}&=&\frac{2\,{\rm e}^{5/6}}{\pi^2\,(N_c-1)!(N_c-2)!}\,{\rm
 e}^{-1.511374\,N_c+0.291746\,n_f}
 \ \ \mbox{(Ref.~\cite{dMS})};  \\ 
 \beta_0&=&\frac{11}{3}N_c-\frac{2}{3}n_f;
 \hspace{1cm} \beta_1=
 \frac{34}{3}N_c^2-\left(\frac{13}{3}N_c-\frac{1}{N_c}\right)\,n_f; 
 \hspace{1cm}
 \gamma_0=3\frac{N_c^2-1}{N_c}.
\end{eqnarray} 
For the case $n_f=0$, relevant for our comparison with lattice results
below, the reduced size distribution $D(\rho )$, Eq.~(\ref{dens}), equals
the proper size distribution, i.e. the number of
instantons per unit volume per unit size, $dn_I/d^4x\,d\rho$, (in the
dilute gas approximation).  
On the other hand, for $n_f>0$,  the proper size distribution reads   
\begin{eqnarray}
\frac{dn_I}{d^4x\,d\rho}\simeq\,
D_m({\rho})= D({\rho})\,\prod_{i=1}^{n_f}(\rho\,m_i(\mu_r)) \,
(\rho\,
\mu_r)^{n_f\,\gamma_0
\frac{\alpha_{\overline{\rm MS}}(\mu_r)}{4\pi}} ,
\label{densm} 
\end{eqnarray}
for small $\rho\,m_i(\mu_r)$, where $m_i(\mu_r)$ are the 
running quark masses.
However, even for $n_f>0$, only the reduced size
distribution $D(\rho )$ appears in cross-sections like Eq.~(\ref{cs}), 
since for small $m_i$, the mass-dependent factor in Eq.~(\ref{densm})
is cancelled by 
corresponding terms from the external quarks.

The powerlaw behaviour of the (reduced) $I$-size distribution,
\begin{equation} 
\label{powerlaw}
D({\rho})\sim \rho^{\beta_0 -5 +\mathcal{O}(\alpha)},
\end{equation}
generically causes the dominant contributions to the $I$-size
integrals to originate from the infrared (IR) regime (large $\rho$) and thus
often spoils the applicability of $I$-perturbation theory.  
In deep-inelastic scattering, however, one parton $q^\prime$ say, carries a
spacelike virtuality, such that  
the contribution of large
instantons to the integrals (\ref{cs}) is exponentially suppressed 
(c.\,f. Eq.~(\ref{formfac})),
\begin{equation}
\label{cutoff}
F(\sqrt{-q^{\prime\,2}}\,\rho)\,
      F(\sqrt{-q^{\prime\,2}}\,\overline{\rho})
\propto {\rm e}^{-Q^\prime\,(\rho+\overline{\rho})};
\hspace{2ex}{\rm for\ }Q^\prime=\sqrt{-q^{\prime\,2}}>0,
\end{equation}
and the predictivity of $I$-perturbation theory is retained for
sufficiently large $Q^\prime$.

A second important ingredient into Eq.~(\ref{cs}) is the function
$\Omega(U,R^2/(\rho\overline{\rho}),\overline{\rho}/\rho)$, appearing
in the exponent with a large numerical 
coefficient $4\pi/\alpha$. It incorporates the effects of final-state
gluons. Within strict $I$-perturbation theory, it is given in form
of a perturbative expansion~\cite{rs-pl,holypert} for large $R^2$, while in
the so-called $\iai$-valley approximation$^{\ref{iai}}$~\cite{yung}
$\Omega$ is associated  with an analytically known 
closed expression\footnote{For the most attractive relative 
$\iai$-orientation the form of the $\iai$-valley action was first given in
Ref.~\cite{valley-most-attr-orient}.
}~\cite{valley-gen-orient} for the interaction between $I$ and $\bar{I}$,   
$\Omega= \alpha/(4\pi)S[A^{\iai}_\mu]-1$,
\begin{equation}
\Omega\left(U(v,z,\ldots),\frac{R^2}{\rho\overline{\rho}},
\frac{\overline{\rho}}{\rho}\right)=\left(\frac{1}{2}-2\,z\right
)\left (1+{v}^{2}   
\right )\Sigma_{{1}}+\left (\left (\frac{1}{2}-2\,z+4\,{z}^{2}
\right )\left (1+{v}^{2}\right )^{2}+2\,{v}^{2}\right )\Sigma_{{2}}
\label{valley}
\end{equation}
with
\begin{equation}
\label{vars}
\begin{array}{lclcl}
\Sigma_{{1}}&=&-2\,{\frac {1-{\lambda}^{4}+4\,{\lambda}^{2}
\ln (\lambda)}{\left ({\lambda}^{2}-1\right )^{3}}};\hspace{9ex}
\Sigma_{{2}}&=&{\frac {1-{\lambda}^{2}+\left (1+{\lambda}^{2}\right )\ln 
(\lambda)}{\left ({\lambda}^{2}-1\right )^{3}}};\\[4ex]
\lambda&=&\frac{1}{2}\,\xi+\frac{1}{2}\,\sqrt {{\xi}^{2}-4}; \hspace{9ex}
\xi&=&\frac{R^2}{\rho\overline{\rho}}+\frac{\overline{\rho}}{\rho}+
\frac{\rho}{\overline{\rho}}\, .
\end{array}
\end{equation}
The $\iai$-valley interaction (\ref{valley}) only depends on two of
the four angle variables necessary in general to parametrize the
relative color orientation $U$ (see e.g. Ref.~\cite{bbgg}). In terms of 
these variabels ($v=\cos\alpha,\ z=\cos(\delta)^2$) and an additional
azimuthal angle variable $\phi$, the integration over the $SU(3)$ group
measure then takes the form~\cite{mrs2} 
\begin{equation}
\int dU=\frac{4}{\pi}\,
\int_{-1}^{1}\!\mid v\mid \left (1-{v}^{4}
\right ){dv}\int_{0}^{\pi }d\phi\!\int_{0}^{\frac{1}{2}\,{
\frac {1-2\,v\cos(\phi)+{v}^{2}}{1+{v}^{2}}}}\!\frac{dz}{1-2\,
v\cos(\phi)+{v}^{2}}=1.
\end{equation}

With these prerequisites in mind, let us next consider
the group-averaged distribution of $\iai$ pairs, 
$d n_{\iai}/d^4 x\,d^4 R\,d\rho\,d\overline{\rho}$. 
This quantity is of basic interest for $I$-induced scattering processes
(\ref{cs}), crucially depends  on the $\iai$-interaction $\Omega$ 
and may be measured on the lattice. For the pure gauge theory\footnote{
For $n_f>0$, there is an additional fermionic interaction factor 
$\mid\int d^4x\, \kappa_{0\,{ I}}^\dagger (x)\,[\ii\not{D}^{ (\iai)}]\,
 \kappa_{0\,{\ai}}(x-R)\mid^{2\,n_f}$ under the group integral in
 Eq.~(\ref{IaI}), with $ \kappa_0$ being the fermionic zero modes.} 
($n_f=0$) and in the dilute-gas approximation, it
takes the form~\cite{rs-pl,mrs2}   
\begin{eqnarray}
\label{IaI}
\frac{d n_{\iai}}{d^4 x\,d^4
R\,d\rho\,d\overline{\rho}} \simeq 
      D_{\iai}(R,\rho,\overline{\rho})= D(\rho )\,
      D(\overline{\rho})
      \ \int dU\, 
      \exp\left[-\frac{4\pi}{\alpha_{\overline{\rm
      MS}}(s_{\iai}/\,\sqrt{\rho\overline{\rho}})}\, 
 \Omega
 \left(U,\frac{R^2}{\rho\overline{\rho}},\frac{\overline{\rho}}{\rho} 
  \right) \right], 
\end{eqnarray}
with the scale factor $s_{\iai}=\mathcal{O}(1)$ parametrizing the
residual scheme dependence. 

In the deep-inelastic regime, relevant for instanton searches at HERA, the
collective coordinate integrals involved in the hard
cross-section (\ref{cs}) are known to be dominated by a unique saddle
point~\cite{rs-pl}.  Therefore, the Bjorken variables
      ($Q^\prime, \xpr$) of the $I$-induced subprocess in $e^\pm P$
      scattering are effectively related in a one-to-one
      correspondence to the collective coordinates as (c.\,f.  also
      Eqs.~(\ref{cs}), (\ref{cutoff})),    
      \begin{equation}
      \label{qual-exp}
      \rho \sim \overline{\rho}\sim 1/Q^\prime \ {\rm and}\  
      R^2\sim 1/(p+q^\prime)^2 \ \Rightarrow \ 
      \rho\overline{\rho}/R^2 \sim 
      (p+q^\prime)^2/Q^{\prime 2}=1/\xpr -1.
      \end{equation}  
Restrictions on the collective coordinates obtained from confronting
predictions of $I$-per\-turbation theory with lattice results may thus
be directly translated into a fiducial kinematical region for the
predicted rate of $I$-induced processes at HERA~\cite{rs-pl}. This
fact illustrates the great importance of lattice simulations for
$I$-induced scattering processes and largely motivates the present analysis. 

{\bf 3.}
Before turning to our concrete analysis, we first
have to address various topical problems associated with lattice data
on instantons. The recent quenched QCD results ($n_f=0$)  from the
UKQCD collaboration~\cite{ukqcd,mike}, which we shall use here,
involve a lattice spacing $a = 0.05 - 0.1$ fm and (roughly constant) volumes  
$V=l_{\rm space}^{\,3}\cdot l_{\rm time}$ of $32^3\, 64\,a^4$ at
$\beta(a)\equiv6/g^2_0(a)=6.4$, $24^3\,  
48\,a^4$ at $\beta=6.2$ and $16^3\,48\,a^4$ at $\beta=6.0$.

In principle, such a lattice allows to study the properties of an ensemble of
topological gauge field fluctuations ($I$'s and $\overline{I}$'s) 
with sizes $a < \rho < V^{1/4}$. However,
in the raw data, ultraviolet fluctuations of wavelength
$\mathcal{O}(a)$ dominate and completely mask the topological effects
which are expected to be of much larger size. 
In order to make the latter  visible, a certain ``cooling'' 
procedure has to be applied first. 
Cooling is a technique for removing the high-frequency non-topological
excitations of the gauge-field, while affecting the
topological fluctuations  of much longer wavelength $\rho \gg a$
comparatively  little. After cooling, $I$'s and $\overline{I}$'s
can clearly be identified as bumps in  the topological
charge density and the Lagrange density. By applying sophisticated
pattern-recognition algorithms, various distributions characterizing
the ensemble of $I$'s and $\overline{I}$'s may be extracted.  

However, during cooling instantons are also
removed; either if they are very small (lattice artifacts) or through 
$I \bar{I}$ annihilation. In Ref.~\cite{ukqcd}, so called
under-relaxed (or slow) cooling is used  to reduce the latter problem. 
Generically, the cooling procedure introduces substantial uncertainties in 
the extraction of physical information on the topological structure of 
the QCD vacuum. No objective criterium is yet available on the
quantitative amount of ``vacuum damage'' introduced by it (see
Refs.~\cite{lattice,ukqcd}).

Since one is interested in the physics of the uncooled vacuum, 
it appears suggestive (at first) to try and extrapolate~\cite{lattice} the 
measured distributions to the limit of a vanishing number of cooling
sweeps, $n_{\rm cool} \to 0$. However, to do so would be to ignore the
fact that the pattern-recognition procedures become increasingly
unreliable near that limit, since  the extracted instanton density appears to
strongly increase for decreasing $n_{\rm cool}$.   
Below a certain minimal number of cools, $(n_{\rm cool})_{\rm min}\,$,
even the total topological charge $Q=n_I-n_{\ai}$ starts to deviate
from the {\it plateau} value it has for a whole range of larger
$n_{\rm cool}$.  

In view of these objections, we follow the alternative strategy 
of Ref.~\cite{ukqcd} to extract information about the physical
vacuum. The method incorporates certain
scaling studies that are important for the continuum limit $a\to 0$
and consists in the following steps: First, at $\beta=6.0$, say, one picks the
smallest number of cooling sweeps  for which the pattern-recognition
starts to become reliable and the cooling history of the total topological
charge enters a plateau.    
For straight cooling, one finds $(n_{\rm cool})_{\rm min}\simeq 10$
corresponding to $\simeq 23$ under-relaxed cools~\cite{mike}.  
The next step is to search for further, so-called {\it equivalent} pairs 
$(\beta,n_{\rm cool})$,
for which shape and normalization of specific distributions
(expressed in physical units\footnote{\label{sommer}For consistency
and minimization
of uncertainties, one should use only a single dimensionful quantity
to relate lattice units and physical units. Throughout our analysis,
all dimensions are therefore  expressed by the so-called Sommer
scale~\cite{sommer,ar0} $r_0$, with $2\,r_0\simeq 1$ fm, which we
prefer over the string tension~\cite{ukqcd}.}) essentially 
remain the same (see also Ref.~\cite{USMT}). The result of
Ref.~\cite{ukqcd} were the pairs 
$(\beta,n_{\rm cool})= (6.0,23),\ (6.2,46),\ (6.4,80)$.
With these equivalent, {\it scaling}
data sets, the continuum limit is then performed with
hopefully little ``vacuum damage'' due to cooling. 

Yet, the crucial question remains at this point, 
whether the choice $(\beta,n_{\rm cool})= (6.0,23)$ as a ``starting
pair'' for the equivalent data is somehow distinguished, i.\,e. {\it
physically} meaningful. In this context, let us present an intriguing
line of arguments:

``Equivalent'' data should first of all be
characterized by a similar amount of cooling. For $n_{\rm cool}$
cooling sweeps,  an effective ``cooling radius'' $r_{\rm cool}$ in
physical units$^{\ref{sommer}}$ may be defined, via a (qualitative) random walk argument, as
\begin{equation}
\frac{r_{\rm cool}}{2\,r_0}={\rm const}\, (\sqrt{n_{\rm cool}})^{1+\delta}\,
\frac{a}{2\,r_0};\hspace{6ex} \delta {\rm\ small}.
\label{rcool}
\end{equation} 
The three equivalent pairs $(\beta,n_{\rm cool})$ above, indeed, turn 
out to have a very similar value of $r_{\rm cool}$ (see Table~\ref{coolrad}). 
\begin{table}[ht]
\begin{center}
\begin{tabular}{|l||c|c|c|}\hline
\rule[-3mm]{0mm}{8mm}($\beta,n_{\rm cool}$) & ($6.0,23$) & ($6.2,46$)
& ($6.4,80$)\\\hline\hline 
\rule[-3mm]{0mm}{8mm}$\sqrt{n_{\rm cool}}\,\frac{a}{2r_0}$&0.447(2) & 
0.460(2) & 0.459(2)\\\hline
\end{tabular}
\end{center}
\caption[dum]{\label{coolrad} The ``cooling radius'' for the
equivalent data sets is similar.}
\end{table}

\begin{figure}[th]
\begin{center}
\epsfig{file=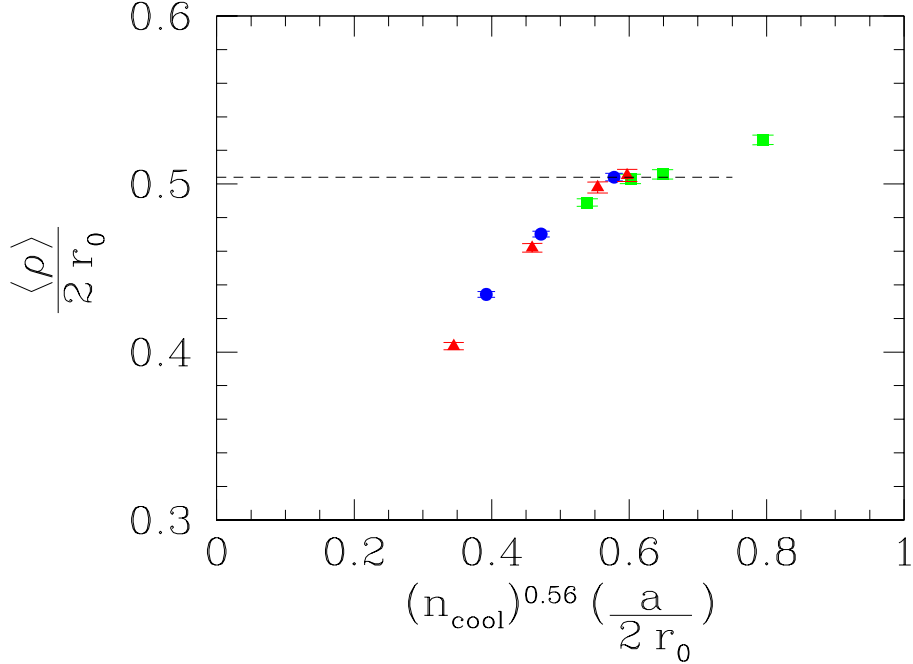,angle=-90,width=12cm}
\end{center}
\caption[dum]{\label{scaling}The average instanton size from
Ref.~\cite{ukqcd} in physical units
for $\beta =6.0\,[\Box ]$, $6.2\,[\circ ]$, $6.4\,[\triangle ]$ and
for {\it all} available values of $n_{\rm cool}$ as a
function of the effective cooling radius (\ref{rcool}). There is a 
clear indication of a {\it plateau} corresponding to $\langle\rho\rangle\simeq
0.5$  fm, which is interpreted in the text. The quantity $r_0$ denotes 
the Sommer scale$^{\ref{sommer}}$ with $2\,r_0\simeq 1$ fm.
}
\end{figure}

Amazingly, the usefulness of the variable (\ref{rcool}) seems to be
much greater. In Fig.~\ref{scaling}, we illustrate that
the average instanton size $\langle\rho\rangle$ in physical units from
Ref.~\cite{ukqcd}, for {\it 
all} available values of ($\beta,n_{\rm cool}$), fits onto a {\it single}
smooth curve as function of $r_{\rm cool}$ in Eq.~(\ref{rcool}) with
$\delta$ small. In contrast, one obtains three widely spaced curves,
if $\langle\rho\rangle/2r_0$ is displayed as a function of $n_{\rm
cool}$ alone for fixed $\beta=6.0,\,6.2$ and $6.4$, respectively.
The same scaling with  $r_{\rm cool}$ holds practically for all quantities
extracted in Ref.~\cite{ukqcd}, characterizing the instanton ensemble on the
lattice.

Perhaps the most striking feature of Fig.~\ref{scaling} is the clear
indication of a {\it plateau} corresponding to $\langle\rho\rangle\simeq
0.5$  fm. Let us try to provide some interpretation for 
it. Suppose, the localized fluctuations on the lattice are a mixture
of two distinct ensembles, separated by
a certain gap in their respective average sizes. For example, one expects 
fluctuations that are lattice artifacts typically located around smaller
sizes $\mathcal{O}({\rm few}\cdot a)$, while the physical instanton sizes
are  peaking around $\rho\simeq r_0\simeq 0.5$ fm. By
gradually increasing the cooling radius $r_{\rm cool}$, we first
filter out the smaller artifacts, leading clearly to a corresponding
increase in the effective $\langle\rho\rangle$. When
the radius of our filter further increases, one then expects a {\it
plateau} in $\langle\rho\rangle$ to
indicate that the small artifacts have been mostly wiped out, while
the much larger physical instantons are not yet strongly
affected. Eventually, the  instantons will also start to be
erased, leading again to an increase in $\langle\rho\rangle$. 

In conclusion, we think that the plateau region 
is the correct regime for extracting physical information about
instantons. The {\it equivalent} data sets of Ref.~\cite{ukqcd} are
precisely located around the plateau.

Our scaling variable $r_{\rm cool}$ allows to combine lattice data for 
a whole range of lattice spacings $a$ and cools $n_{\rm cool}$, which
strongly helps to increase the separation power between the $a$-dependent
artifacts and the instantons of a certain {\it physical} size.
In future analyses, it may be worthwhile to determine the unknown constant in
the definition~(\ref{rcool}) of $r_{\rm cool}$, e.\,g. by means of
calibration via cooling an isolated instanton until it disappears.

{\bf 4.} We are now ready to discuss the continuum limit of the UKQCD data
for the size distribution of $I$'s and $\ai$'s, $dn_{I+\ai}/d^4x\,d\rho$, 
and to compare it with the dilute-gas prediction
$D_{I+\ai}(\rho )=2\,D(\rho)$ of $I$-perturbation theory,
Eq.~(\ref{dens}), for $n_f=0$. 

\begin{figure} [ht] 
\begin{center}
\parbox{10.55cm}{\vspace{-0.25cm}\epsfig{file=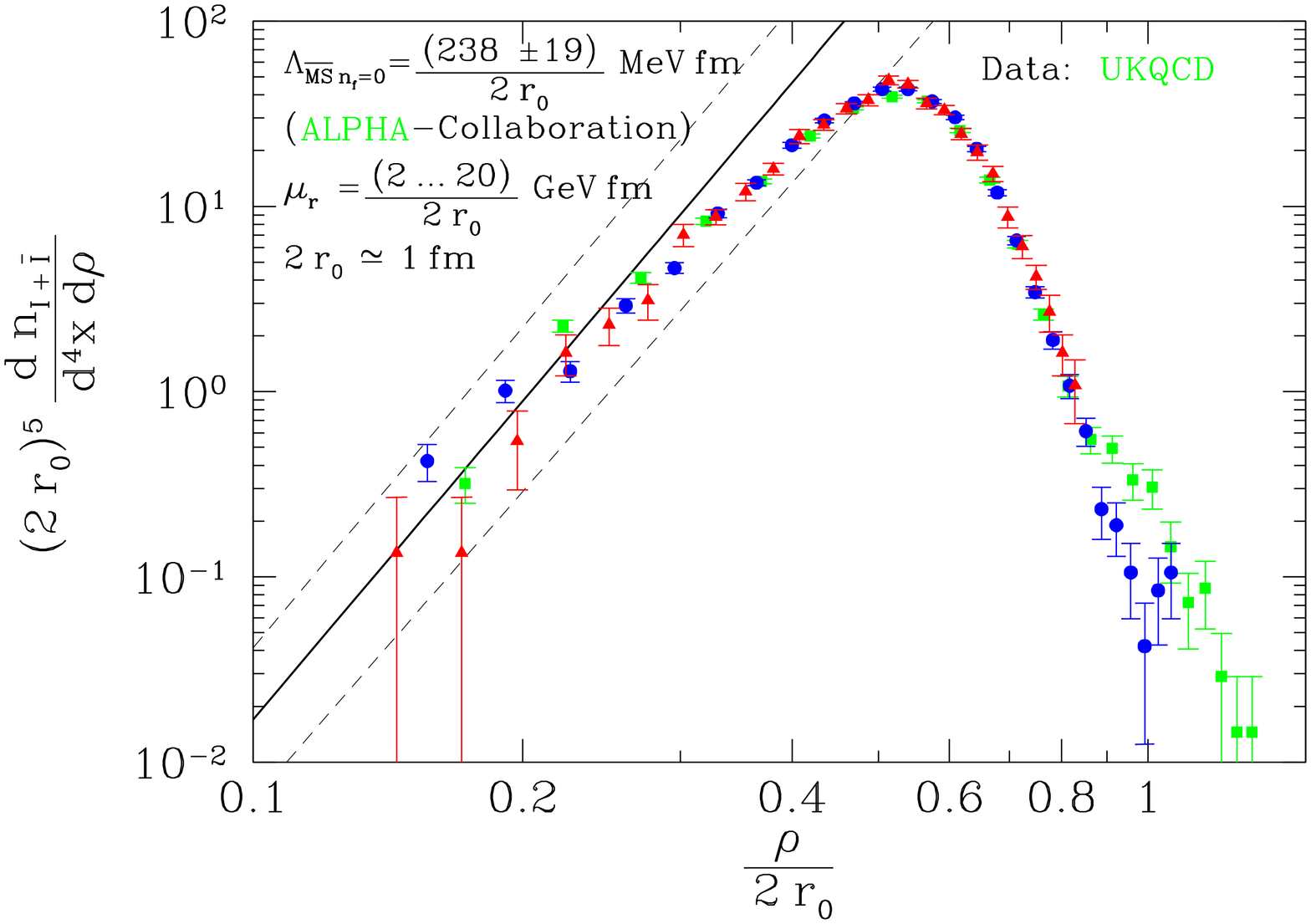,%
angle=-90,width=11.05cm}}
\parbox{6.15cm}{\vspace{-0.3cm}\epsfig{file=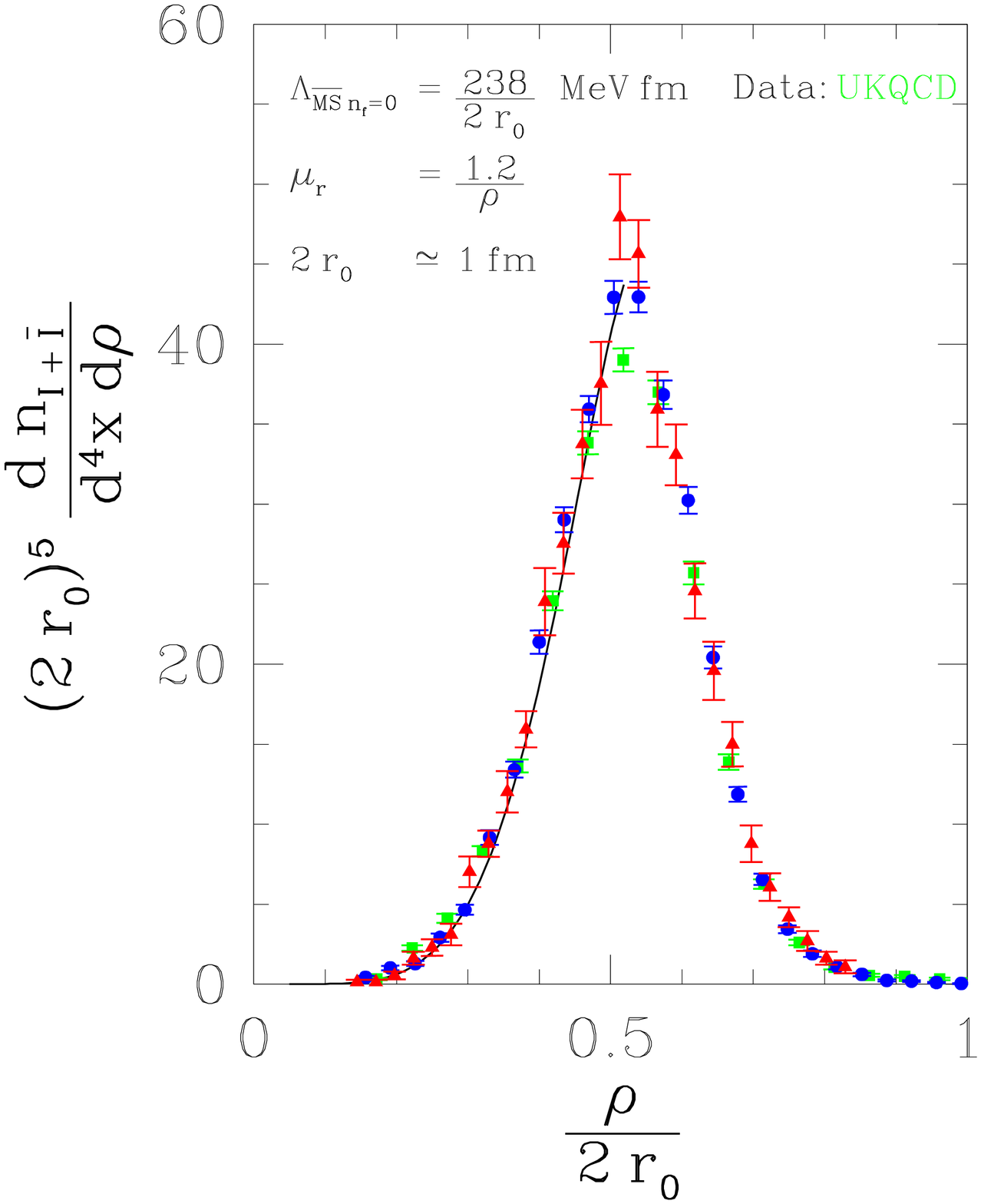,width=6.15cm}}
\end{center}
\caption[dum]{\label{lattice}
    Continuum limit of ``equivalent'' UKQCD
    data~\cite{ukqcd} for the 
    ($I+\overline{I}$)-size distribution at $(\beta,n_{\rm
    cools})=(6.0,23)\,[\Box ],\ (6.2,46)\,[\circ ],\
    (6.4,80)\,[\triangle ]$. The striking agreement with $2\,D(\rho)$      
    of $I$-perturbation theory from (\ref{dens}) is
    apparent. The 3-loop form of 
    $\alpha_{\overline{\rm MS}}$ with $\Lambda_{\overline{\rm MS}\,n_f=0}$ from
    ALPHA~\cite{alpha} was used.  
    (a) Log-log plot to exhibit the expected power law $\sim \rho^6$
        and the agreement in magnitude for small $\rho$ over a wide
        range of $\mu_r$.  The dashed error band results from varying  
    $\Lambda_{\overline{\rm MS}\,n_f=0}$ and $\mu_r$ within its error and
    given range, respectively;    
    (b) For $\mu_r=1.2/\rho$, the agreement extends up to the peak.}
\end{figure}

We note, first of all, that  the (small) residual $a$ dependence of the {\it
equivalent} data for the size distribution in physical units may be
parametrized as
\begin{equation}
(2\,r_0)^5\,\frac{dn_{I+\ai}}{d^4x\,d\rho}= \mbox{\
function}\,\left(\frac{\rho}{\langle\rho(a)\rangle}\right).
\label{shapedist}
\end{equation}   
The continuum limit of Eq.~(\ref{shapedist}) can thus be performed
quite reliably, by simply rescaling the arguments 
$\rho \to \langle\rho(0)\rangle/\langle\rho(a)\rangle\cdot \rho$. Here, 
$\langle\rho(0)\rangle/2r_0 =0.518(5)$ denotes the continuum limit of the
weakly varying average $\rho$ values, $\langle\rho(a)\rangle$. 
As in Ref.~\cite{ukqcd}, $\langle\rho(a)\rangle/2r_0$ was extrapolated
linearly in $(a/r_0)^2$. The resulting continuum size distribution
obtained by the above rescaling from the three equivalent data sets is
displayed in Fig.~\ref{lattice}. It  scales nicely. 

We are now ready to perform a quantitative comparison with 
$I$-perturbation theory. The corresponding prediction     
(\ref{dens}) is a power law
for small $\rho$, i.\,e. approximately $D \sim \rho^6$ for $n_f=0$. Due 
to its 2-loop RG-invariance the normalization of
$D_{I+\overline{I}}(\rho)$ is practically
independent of the renormalization scale $\mu_r$ over a wide range.
It is strongly dependent on
$r_0\,\Lambda_{\overline{\rm MS}}$, for which we take the most recent,
accurate lattice result by the ALPHA collaboration~\cite{alpha},
$ 2\,r_0\,\Lambda_{\overline{\rm MS}\,n_f=0}=(238\pm 19)$ MeV\,fm.
In Fig.~\ref{lattice}~(a) we display both this parameter-free
prediction (\ref{dens}) of $I$-perturbation theory  and the continuum
limit of the  UKQCD data  in a log-log
plot, to clearly exhibit the expected power law in $\rho$. The
agreement in shape\footnote{The agreement in  shape up to $\rho\simeq
0.3$ fm has been noted by us
already previously~\cite{rs-pl}. Here, we have extended the analysis to
the continuum limit and, in particular,
verified the agreement in normalization.} {\it and} normalization for
$\rho/2r_0\lwig 0.3$  is striking, indeed, 
notably in view of the often criticized cooling procedure and the 
strong sensitivity to  $\Lambda_{\overline{\rm MS}}$.
One may argue that a choice $\rho\mu_r=\mathcal{O}(1)$ is theoretically
favoured, since it tends to suppress $\ln (\rho\mu_r)$ corrections. Indeed, as
apparent from Fig.~\ref{lattice}~(b), the agreement extends in this 
case up to the peak around $\rho/2r_0\simeq 0.5$.

At this point, an interesting way to proceed is to define a
non-perturbative ``instanton-scheme'' for $\alpha$ by the
requirement that the theoretical expression (\ref{dens}) for the size
distribution $D_{I+\overline{I}}(\rho)$ {\it precisely} reproduces 
the corresponding lattice data,
\begin{equation} 
\rho^5\,\frac{dn_{I+\ai}}{d^4x\,d\rho}\equiv  
2\,d_{\overline{\rm MS}}
\left(\frac{2\pi}{\alpha_{\rm I}(\frac{1}{\rho})}\right)^{2\,N_c} 
\exp{\left(-\frac{2\pi}{\alpha_{\rm I}(\frac{1}{\rho})}\right)}s_{\rm
I}^{\beta_0+\frac{\alpha_{\rm
I}(\frac{1}{\rho})}{4\pi}(\beta_1-4\,N_c\beta_0)}. 
\label{ischemeexp}
\end{equation} 
By comparing in the perturbative regime with the $\overline{\rm MS}$
form (\ref{dens}) for $\mu_r=s/\rho$, we arrive at a standard scheme
conversion formula, 
\begin{equation}
\alpha_{\overline{\rm MS}}(s/\rho)=\alpha_{\rm I}(1/\rho)
-\frac{\beta_0}{2\pi}\ln (s/s_{\rm I})\,\alpha_{\rm I}(1/\rho)^2
+\mathcal{O}(\alpha_{\rm I}(1/\rho)^3),
\end{equation}
with 
\begin{equation}
s_{\rm I} = \frac{\Lambda_{\overline{\rm MS}}}{\Lambda_{\rm
I}}\hspace{1cm} \mbox{\ and\ }\hspace{1cm} \alpha_{\overline{\rm MS}}(s_{\rm
I}/\rho)\simeq\alpha_{\rm I}(1/\rho). 
\label{ischeme}
\end{equation} 
Next, we note that as a function of $\alpha_{\rm I}$,
$D_{I+\overline{I}}(\rho)$ has a peak structure like the lattice data, 
with maximum at
\begin{equation}
\alpha_{\rm I\,\mid
max}=\frac{\pi}{N_c}\,\frac{2}{1+\sqrt{1+\frac{1}{N_c}\ln(s_{\rm 
I})(2\beta_0-\frac{\beta_1}{2 N_c})}}. 
\end{equation}
The scale factor $s_{\rm I}$ of the $I$-scheme is non-perturbatively
and uniquely determined by the peak value of the lattice data for
the size distribution. Using the very good Gaussian fit to all data
in Fig.~\ref{lattice}~(b),
\begin{equation}
\label{D-fit}
(2\,r_0)^5\,\frac{dn_{I+\ai}}{d^4x\,d\rho}_{\mid {\rm fit}}=
\frac{(a\rho )^6 \exp\left(-(c\rho )^2\right)}{(1+(b\rho )^p)} ;
\hspace{6ex}
\left\{
\begin{array}{lcl}
a  & = &4.908(40)/2r_0 \\
b  & = &1.683(15)/2r_0 \\
c  & = &2.525(52)/2r_0 \\
p      & = &12.43(25) 
\end{array}
\right.
\end{equation}
we find 
\begin{equation}
s_{\rm I}= 1.18 \mbox{\ and\ } \alpha_{\rm I\,\mid max}=0.98.
\end{equation} 
In the perturbative regime, the $I$-scheme is thus quite close to the
$\overline{\rm MS}$-scheme (c.\,f. Eq.~(\ref{ischeme})). 
\begin{figure} [th]
\vspace{-0.4cm}
\begin{center}
\parbox{8.1cm}{\epsfig{file=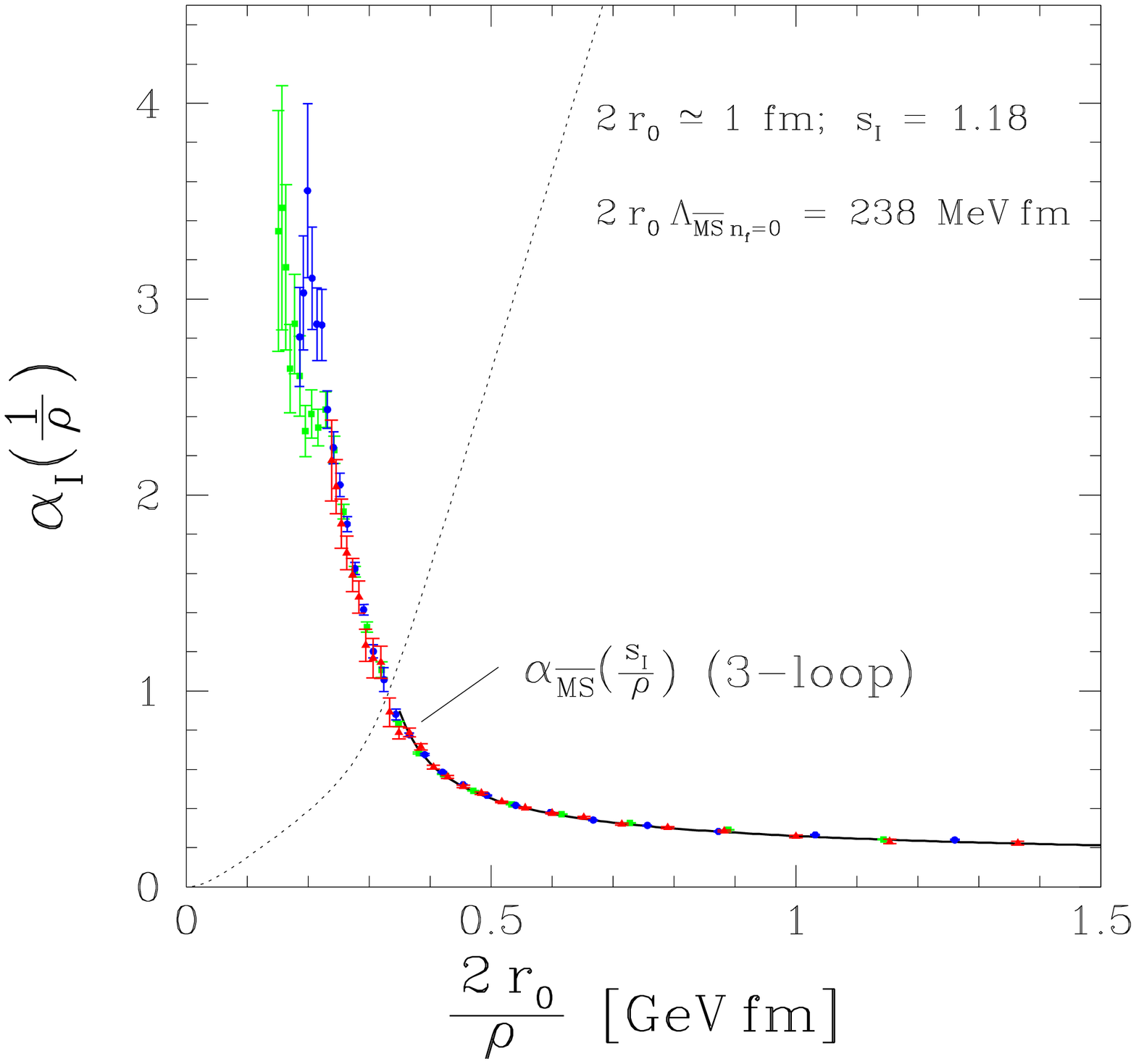,angle=-90,width=8.1cm,clip=}}
\parbox{8.7cm}{\epsfig{file=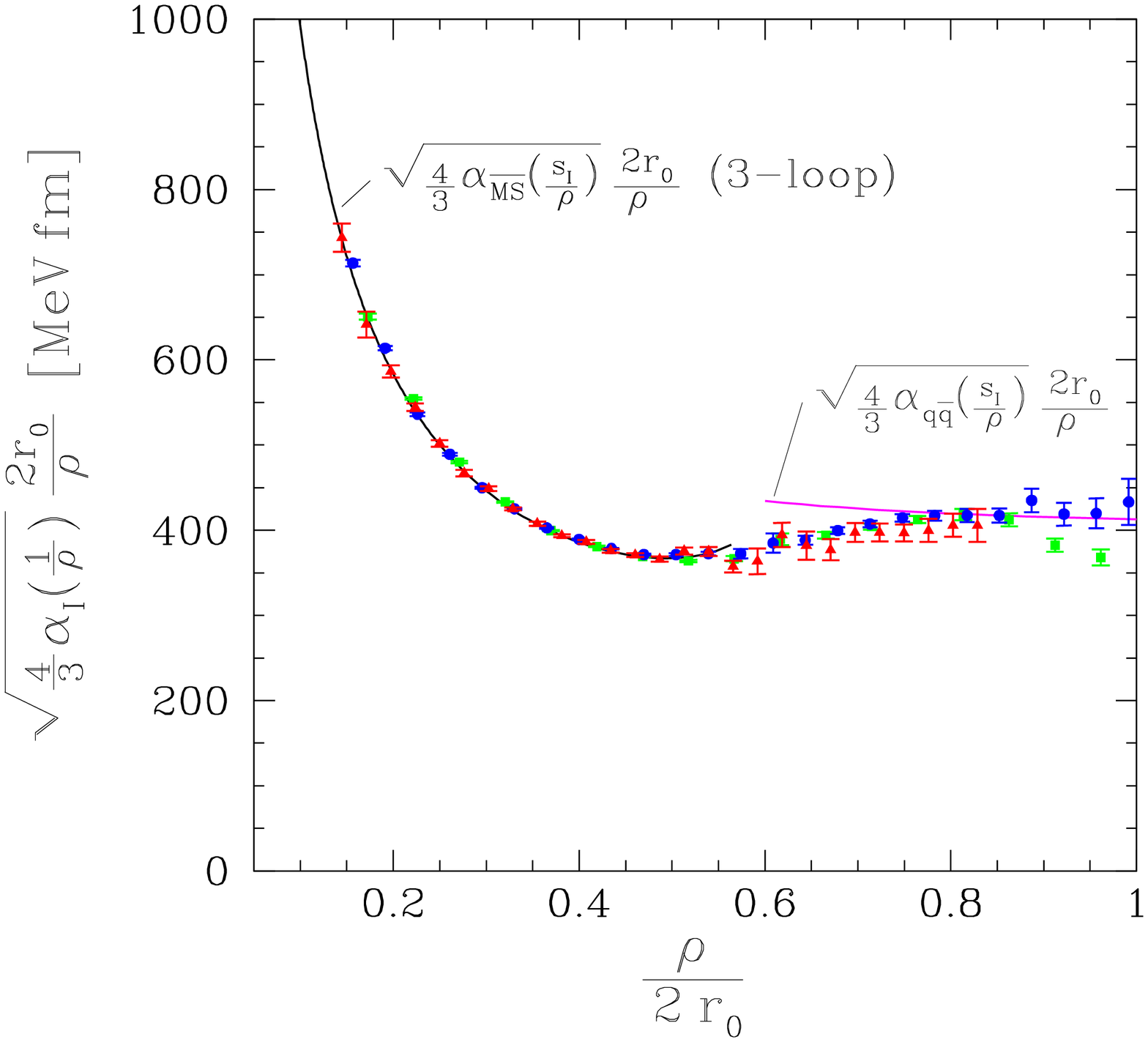,angle=-90,width=8.7cm}}
\caption[dum]{\label{alphai} (a) Running coupling $\alpha_{\rm I}$ in the
``instanton'' scheme, extracted via Eq.~(\ref{ischemeexp}) from the $I +\ai$
size distribution. The same UKQCD data as in Fig.~\ref{lattice} have 
been used. The dotted line denotes a possible second solution vanishing in the
IR limit. The excellent agreement with the corresponding 3-loop
perturbative $\alpha_{\overline{\rm MS}}(s_{\rm
I}/\rho)\simeq\alpha_{\rm I}(1/\rho)$ down to $1/\rho\simeq 0.4$ GeV
is apparent (solid line). (b) At large distances, $\alpha_{\rm I}$ from
(a) appears to be quite close to $\alpha_{\rm q\overline{q}}$ defined
    via the static $q\overline{q}$ potential, with $(4/3r^2)\,\alpha_{\rm
q\overline{q}}(1/r)$ approaching the string tension $\sigma$ for
large $r$ and~\cite{ar0} $2\,r_0\,\sqrt{\sigma}=(472.4\pm 4.3)$
MeV\,fm.}
\end{center}
\end{figure}

We are now able to extract for each data point of the $I+\ai$ size
distribution a corresponding value of $\alpha_{\rm I}$, as illustrated 
in Fig.~\ref{alphai}. Actually, besides the asymptotically free
solution [$\alpha_{\rm I}$] to Eq.~(\ref{ischemeexp}), there is a second one [$
\alpha^{(2)}_{\rm I}$] (dotted line in Fig.~\ref{alphai}),  with 
a monopole-like relation to $\alpha_{\rm I}$,  
\begin{equation}
    \alpha^{(2)}_{\rm I}\cdot \alpha_{\rm I}=\mathcal{O}(1), 
\end{equation}
vanishing in the IR-limit and increasing like
$\alpha^{(2)}_I(\mu)\propto (\mu/\Lambda)^{\beta0/2N_c}$ in the
UV-limit. This observation may be of interest in the context of
Ref.~\cite{alpha2}.   
  
It is remarkable that in the non-perturbative long-distance regime,
$\rho/2r_0\gwig 0.5$, $\alpha_{\rm I}(1/\rho)$ from the $I +\ai$ size
distribution tends to approach $\alpha_{\rm
q\overline{q}}(1/\rho)$ defined via the static $q\overline{q}$ potential, 
$V_{\rm q\overline{q}}$,
\begin{equation}
 \frac{4}{3}\frac{\alpha_{\rm q\overline{q}}(1/r)}{r^2}\equiv \frac{d V_{\rm
 q\overline{q}}(r)}{dr}
 \simeq\ \sigma + \frac{\pi}{12}\frac{1}{r^2} 
\stackrel{r\rightarrow \infty}{\Rightarrow} \sigma, 
\end{equation}  
with string tension $\sigma$. This is illustrated in Fig.~\ref{alphai}~(b).

In summary, $\alpha_{\rm I}$ as non-perturbatively defined through
Eq.~(\ref{ischemeexp}), is close to the $\overline{\rm MS}$-scheme in
the perturbative short-distance region and to the $q\overline{q}$-scheme in the
non-perturbative long-distance regime. Figs.~\ref{lattice}~(b)
and~\ref{alphai}~(b) illustrate that the transition between  
perturbative and confinement-sensitive physics seems to be rather sharp
and located where the instanton density peaks, around $\rho\simeq 0.5$ fm. 

The  above results also suggest the following working
hypothesis~\cite{durham}, which might considerably enlarge the
region of applicability of predictions like (\ref{cs}) and (\ref{IaI}) for
observables associated with sufficiently dilute instantons. 
Whenever one encounters  the $I$-size distribution $D(\rho)$ within the
perturbative calculus, one may try to replace the  indefinitely growing 
perturbative expression  $D(\rho)\sim\rho^{\beta_0-5}$ 
for the $I$-size distribution by the functional form 
actually extracted from the lattice (see e.g. Eq.~(\ref{D-fit})). 
Obviously, such an attempt eliminitates the problematic IR divergencies
of $I$-size integrals.  
Due to the strong peaking of the measured size
distribution~(\ref{D-fit}), only $I$'s of size $\rho\simeq0.5$ fm
effectively contribute to the observables of interest\footnote{This philosophy 
is partly also inherent in the $I$-liquid model~\cite{ssh}.} and the
dilute gas approximation may well continue to hold qualitatively up to the 
peak around $0.5$ fm.   

{\bf 5.}
As an immediate first application of this strategy, let us
turn to the group-averaged distribution of $\iai$-pairs,
$D_{\iai}(R,\rho,\overline{\rho})$, in the dilute gas approximation,
Eq.~(\ref{IaI}).  While the conservative prediction of
instanton perturbation theory is restricted to $\rho,\overline{\rho}\lwig
0.3 - 0.5$ fm, UKQCD lattice results~\cite{ukqcd,mike} 
are available up to now only for the $\iai$-distance distribution, integrated 
over {\it all} values of $\rho,\overline{\rho}$,    
\begin{equation}
\label{IaIdist}
\frac{d n_{\iai}}{d^4 x\,d^4 R}=
\int\,\int d\rho\,d\overline{\rho}\,\frac{d n_{\iai}}{d^4 x\,d^4
R\,d\rho\,d\overline{\rho}}\,.
\end{equation}

This is the point to exploit the above working hypothesis:
We now replace in the theoretical expression for the
$\iai$-distance distribution (\ref{IaIdist}) along with
Eq.~(\ref{IaI}) the perturbative form for $D(\rho)$ by the Gaussian fit
(\ref{D-fit}) to the UKQCD lattice data. We are left with only one free
parameter,  the scale factor $s_{\iai}=\mathcal{O}(1)$ in the running coupling
$\alpha_{\overline{\rm MS}}(s_{\iai}/\sqrt{\rho\overline{\rho}})$.
On account of the peaking of the size distributions in Eq.~(\ref{IaI}), the
running coupling at a momentum scale around $\mathcal{O}(0.3)$ GeV
enters significantly. In this region, we make use of
the non-perturbatively extracted coupling in the $I$-scheme
(Fig.~\ref{alphai}), with
\begin{equation}
\label{alphams}
\alpha_{\overline{\rm
MS}}\left(\frac{s_{\iai}}{\sqrt{\rho\overline{\rho}}}\right)\Rightarrow 
\alpha_{I}\left(\frac{s_{\iai}}{s_{I}}\frac{1}{\sqrt{\rho\overline{\rho}}} 
\right) ;\hspace{6ex} s_I = 1.18 ,
\end{equation}
as a smooth continuation of the perturbative $\overline{\rm MS}$
coupling (c.\,f. Eq.~(\ref{ischeme})).
In the quantitative evaluation of Eqs.~(\ref{IaIdist}) and~(\ref{IaI}),
we perform two integrals of the group integration and the two
integrals over the $I$- and $\ai$-sizes numerically.  

\begin{figure}[tb]
\begin{center}
\epsfig{file=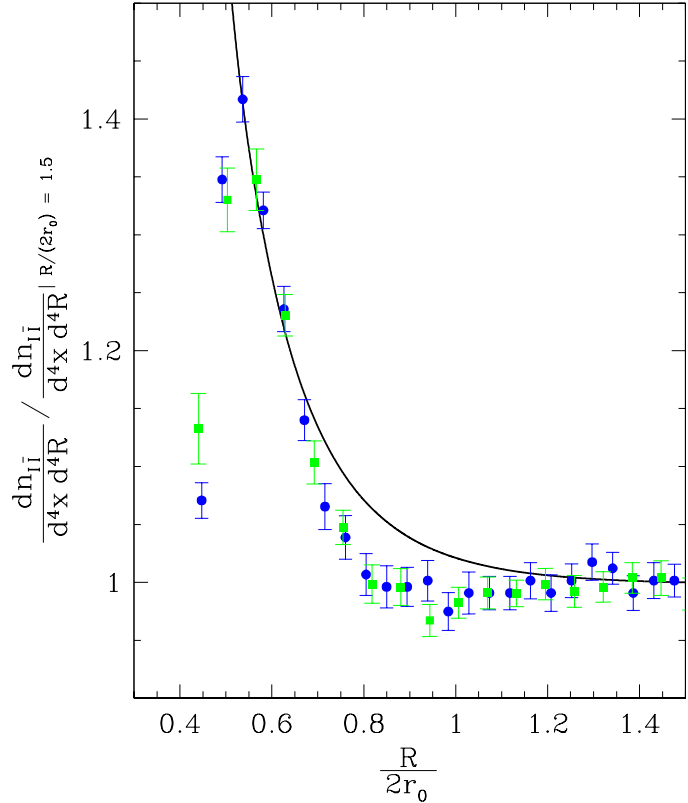,width=9cm}
\end{center}
\caption[dum]{\label{pairdist} Continuum limit of ``equivalent'' UKQCD 
data~\cite{mike} for the $\iai$-distance distribution, 
$dn_{\iai} /d^4x\,d^4 R$, normalized to its value at $R/2r_0=1.5$, at 
$(\beta,n_{\rm
    cools})=(6.0,23)\,[\Box ],\ (6.2,46)\,[\circ ]$. 
As discussed in the text, the solid line denotes the theoretical
prediction from instanton-perturbation theory based on the valley form of
the $\iai$-interaction.  
  }
\end{figure}

Before being able to compare this theoretical prediction of the $\iai$-distance
distribution with lattice results,  
we have to perform the continuum limit of the respective UKQCD
data~\cite{mike}. This we do in complete analogy to the method
described in Sect. 4., Eq.~(\ref{shapedist}), for the $I$-size
distribution. Unfortunately, the equivalent data for $\beta=6.4$ are
missing in this case~\cite{mike}. The equivalent data
for the remaining two $\beta$ values, $(\beta,n_{\rm cool})=(6.0,23)$
and $(6.2,46)$, again are well compatible with a scaling law, like
Eq.~(\ref{shapedist}),
\begin{equation}
\frac{d n_{\iai}}{d^4 x\,d^4 R}\, \Bigg / 
     \frac{d n_{\iai}}{d^4 x\,d^4 R}_{\mid\,R\ {\rm large}}
=\ {\rm function} \left(\frac{R}{r(a)}\right).
\label{Rscaling}
\end{equation}
By extrapolating the scaling factor $\lambda(a)\equiv r(a)/r_{\mid\,
\beta=6.2}$ linearly in $(a/r_0)^2$, we find 
from requiring best overall scaling (\ref{Rscaling}) of the two
equivalent data sets, $\lambda_{\iai}(a)=1.32-69.75(a/2r_0)^2$. 
The resulting ``continuum data'' obtained from the simple rescaling of the 
arguments, $R \to r(0)/r(a)\,R=\lambda(0)/\lambda(a)\,R$, are
displayed in Fig.~\ref{pairdist} and exhibit good scaling.

The solid line in Fig.~\ref{pairdist} denotes the theoretical
prediction discussed above with the scale factor $s_{\iai}=0.82$ being
$\mathcal{O}(1)$ as expected.

We note a very good agreement with the lattice data  down to 
$\iai$-distances $R/\langle\rho\rangle\simeq 1$, corresponding to   
$\xi\gwig 3$ (c.\,f. Eq.~(\ref{vars})).  These
results imply first direct support for the validity of the valley
interaction (\ref{valley}) between instantons and anti-instantons.   

\vspace{0.2cm}
\section*{Acknowledgements}
We thank Hans Joos for constant encouragement and many inspiring
discussions on instantons. To Martin L{\"u}scher we owe constructive
criticism and helpful discussions on the results of the ALPHA
collaboration. Moreover, we also acknowledge discussions with Fritz
Gutbrod, John Negele and Gerrit Schierholz, notably on cooling. Last
not least, we wish to thank Mike Teper  for many helpful communications and
suggestions, as well as for providing us with partly unpublished data
from his work~\cite{ukqcd}.

\end{document}